\documentclass[%
reprint,
superscriptaddress,
 amsmath,amssymb,
 aps,
prb,
raggedbottom,
]{revtex4-2}
\usepackage[linktocpage,colorlinks,breaklinks,linkcolor=blue,citecolor=blue,urlcolor=blue]{hyperref}
\usepackage{graphicx}% Include figure files
\usepackage{dcolumn}% Align table columns on decimal point
\usepackage{bm}% bold math
\usepackage{tabularx}
\usepackage{float}
\usepackage{placeins}

\begin{document}

% \preprint{APS/123-QED}

\title{Calculation of the energies of the multideterminant states of the nitrogen vacancy center in diamond with quantum Monte Carlo}% Force line breaks with \\

\author{K. A. Simula}

\author{I. Makkonen}%
\affiliation{%
  Department of Physics, P. O. Box 43, FI-00014 University of Helsinki, Finland
}%

\date{\today}

\begin{abstract}

Certain point defects in solids can efficiently be used as qubits for applications in quantum technology. They have spin states that are initializable, readable, robust, and can be manipulated optically. New theoretical methods are needed to find the best host materials and defect configurations. Most methods proposed so far rely either on cluster models or restrict the many-body treatment of the defects to a subspace of single-particle orbitals. We explore best practices and theory for the use of quantum Monte Carlo to predict the excitation spectra for spin defects, by using the negatively charged nitrogen vacancy (NV$^-$) center in diamond as a test system. Quantum Monte Carlo can be used to explicitly simulate electronic correlations with larger systems and sets of orbitals than previous methods due to favourable scaling with respect to system size and computing power. We consider different trial wave functions for variational and diffusion Monte Carlo methods, explore the nodal surface errors of the ground and excited state wave functions and study whether the variational principle holds for the excited states. We compute the vertical excitation energies in different simulation cell sizes and extrapolate to infinite system size, and include backflow corrections to the extrapolated energies. The final results for vertical excitation energies are found to overestimate the experimental estimates, but the triplet-to-triplet and singlet-to-singlet transitions are accurate against experiment. Finally, we list further developments for QMC needed to address the problem of accurately predicting structural and spin properties of the solid-state defects. 

\end{abstract}

\pacs{Valid PACS appear here}% PACS, the Physics and Astronomy
\maketitle

\section{Introduction}

Color centers semiconductors have attracted wide interest during the last two decades due to their potential use in many areas of quantum technology. Point defects are competitive candidates as the building blocks of future devices in quantum sensing \cite{schirhagl2014}, networks \cite{nguyen2019}, and computing \cite{weber2010,zhang2020,abobeih2022}. There are also point defect applications for single-photon emitters \cite{aharonovich2016,aharonovich2011} and quantum memory platforms \cite{blencowe2010,bradley2019,wrachtrup2006}

Solid-state qubits can have long coherence times \cite{seo2016}, even up to one minute \cite{bradley2019}, and optical manipulation of the ground-state spin of the defects is possible also at room temperature \cite{dutt2007,koehl2011}. The initialization of the ground-state spin can be achieved by optical pumping \cite{doherty2013}, and optical or electronic readout of the spin state is possible \cite{hegde2020,siyushev2019}. The coupling of the electron spin to nearby nuclear spins can be controlled and monitored \cite{childress2006}, and the resulting multiqubit ensemble enables error correction \cite{waldherr2014}, longer coherence times \cite{seo2016}, the use of nuclear spins as quantum registers \cite{dutt2007}, and implementations of quantum algorithms with a single defect \cite{zhang2020}. The NV$^-$-center in diamond is the best known point defect qubit, but other defects in diamond have also shown promising features for quantum applications \cite{iwasaki2015,sukachev2017}. Silicon and silicon-carbide-based solid-state qubits show potential with industrially scalable materials \cite{koehl2011,christle2015,awschalom2018,bourassa2020}. 

Although many known solid-state systems show promising features for quantum applications, even better systems are very probably yet to to be found. The possibilities for tuning the properties of the solid-state defects for particular applications are only limited by the capability to find, simulate, produce and control the defect systems of various host materials.

Theoretical modeling of the solid-state qubits is important in understanding the properties of known point defects and finding novel defects \cite{dreyer2018}. Defect formation energies and their interaction with other defects and surfaces can be computed with well-established computational machinery, but the determination of ionization energies and magneto-optical properties is much more challenging \cite{gali2019}. The highly correlated defect states couple to the electronic orbitals of the surrounding bulk, creating the need to model the defect and its surroundings accurately.

Density functional theory (DFT) can find accurate atomic structures and properties such as Stokes shifts, and describe excitations within the constrained-DFT approach between single-determinant states \cite{gali2009}. It can also treat simulation cells up to thousands of atoms without unbearable computational cost \cite{thiering2017}. The highly correlated singlet states with multideterminant wave functions \cite{cohen2008}, which are often very important in initialization of a solid-state qubit, can be studied with DFT with a recent implementation \cite{ivanov2023}.

Wave function-based methods can be used to model highly correlated many-body states, but tend to have unfavorable scaling with respect to system size. One way to circumvent this is to use cluster models \cite{delaney2010,bhandari2021}, with cluster size sufficiently small for simulations. An increasingly popular approach is to use embedding schemes, where the bulk solid is modeled with lesser accuracy by using e.g. DFT or Green's function based single-particle methods, and the impurity region and electronic states localized around it are subsequently treated with a wave function based approach \cite{bockstedte2018,sun2016,ma2021}. The embedding schemes have proven very accurate in some cases, but suffer from problems of double counting (although there exist very efficient double counting correction schemes \cite{sheng2022}) and problems in accounting for the screening of the defect states by the electrons in the surrounding bulk of the material. Also the choice of the underlying correlation functional in the DFT calculations can alter the results \cite{muechler2022}.

Quantum Monte Carlo (QMC), particularly the variational and diffusion Monte Carlo (VMC and DMC) flavours, can describe periodic many-body systems including an impurity without separate treatment of bulk and defect states. Especially DMC is accurate in estimating energy of various real systems, and it scales better than most quantum chemistry methods against systems size. It also scales almost linearly with the amount of computing power, a remarkable asset in the age where petascale supercomputers are readily available \cite{foulkes2001}. Another strong feature of QMC is that it is a variational method, enabling systematic improvement of the wave functions describing the many-body states, although there are some limitations of the variational principle for the excited states \cite{foulkes1999}.

Defect formation and migration energies have been computed in various works with DMC\cite{hood2003,alfe2005,parker2011,hood2012}. Optical excitations are seldom studied so far with DMC, but a few results exist \cite{hood2003,ertekin2013,saritas2018}. The study of Hood \textit{et al.} \cite{hood2003} has modeled multideterminant excitation energies with DMC in solid-state defects. Recent study used full configuration interaction QMC calculation to treat an embedded C$_3$N$^-$ region in a cluster of $42$ C atoms \cite{chen2023}, finding strong multiconfigurational nature for the ground and excited states. DMC studies of multideterminant states in molecules are more common \cite{morales2012,rao2022}. Successful periodic DMC simulation of lattice defect states with multideterminant character would enable fully correlated studies of point defect qubits within a larger volume than accessible with current embedding schemes using quantum chemistry methods. This could enable more accurate analysis of the coupling of the defect states to the bulk of the material and smaller finite-size errors. 

Here, we model the many-body states of the NV center with DMC and VMC-optimized Jastrow factors and backflow functions. In Sec. \ref{section: many-body states} we present the relevant many-body states in the NV$^-$-center. Section \ref{section: QMC simulation of the excitations} presents the theory and concepts needed in the study of excitations, and discusses finite size effects and nodal surface errors in DMC simulations. We present results from our computations in Sec. \ref{section: results}, and discuss the significance of the results and possible continuations of the present study in Sec. \ref{section: discussion}.

\section{Many-body states of the NV$^{-}$-center}\label{section: many-body states}

\begin{figure}
  \includegraphics[scale=.45]{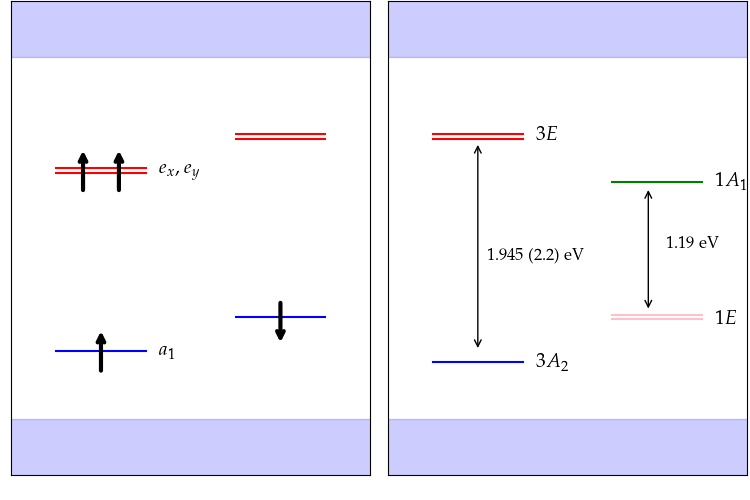}
  \caption{\label{figure: sketch of the multiplet structure} Sketch of energy structure of the single-particle orbitals (left) and the many-body states (right) of the NV$^-$-center. A split between the spin channels of the single-particle orbitals is shown, so that the spin-majority (spin-up) channel in the left has orbitals with lower energies than the spin-minority (spin-down) channel. The electronic occupation of the single-particle orbitals in the $3A_2$ ground state is denoted for spin-up (upward arrow) and spin-down (downward arrow) electrons. The many-body states in energetic order from lowest to highest energy are $3A_2$ (blue), $1E$ (pink), $1A_1 $ (green), and 3E (red). The double arrows denote the triplet-to-triplet and singlet-to-singlet transitions, with experimental values shown in eV. The energies of the doubly degenerate orbitals $e_{x,y}$ and the many-body states $3E$ and $1E$ are represented with two lines, displaced vertically for easier depiction.}
\end{figure}

According to various experimental and theoretical works \cite{davies1976,rogers2008,gali2019,jin2022}, NV$^{-}$-center is known to have $4$ many-body states relevant for the optical cycle of qubit operation, energetically in the following order: $3A_2$, $1E$, $1A_1$, and $3E$. A sketch of the multiplet structure is shown in Fig. \ref{figure: sketch of the multiplet structure}. The Mulliken notation reveals that there is a fully symmetric triplet ground state $3A_2$, a doubly degenerate triplet excited state 3E, and two singlet states between the triplets, fully symmetric $1A_1$ and doubly degenerate $1E$. The total spin $S$ is zero for the singlet states, and $1$ for the triplet states. The projection of the total spin along the symmetry axis of the NV$^-$-center, $S_z$, can be $0$ or $\pm 1$ for the triplet states. 

There are two optical transitions that can be measured accurately: the triplet-to-triplet $3A_2 \leftrightarrow 3E$ and singlet-to-singlet $1E \leftrightarrow 1A_1$. The former has the zero-phonon line of $1.945$ eV \cite{davies1976} and a vertical excitation energy of $2.2$ eV in the ground-state structure, and the latter a zero-phonon line of $1.19$ eV \cite{rogers2008}.  The Stokes and anti-Stokes shifts between the triplet states are known from experiments to be $2.235$ eV and $1.85$ eV, respectively \cite{davies1976}. The transition between the triplet and singlet states can happen via intersystem crossing \cite{thiering2018,gali2019}. These transitions are dark, and the energetic ordering of the singlets in relation to the triplet states is difficult to measure. According to a number of experimental studies \cite{davies1976,rogers2008,goldman2015,goldman2015b} $1E$ lies $0.3-0.5$ eV, and $1A_1$ $1.5-1.7$ eV above the $3A_2$ state.

\section{QMC simulation of the excitations}\label{section: QMC simulation of the excitations}

\subsection{The single-and many-body states of the NV$^-$-center}

The description of the defect system is determined by the Hamiltonian. Within the Born-Oppenheimer approximation, the Hamiltonian for an $N$-electron system is
\begin{align}
  \label{equation: interacting Hamiltonian}
  \begin{aligned}
    \hat{H}=&\sum_{i=1}^N \left(-\frac{1}{2} \bigtriangledown_i^2 + v(\mathbf{r}_i) \right) + \sum_{i=1}^N\sum_{j>i}^N \frac{1}{\left|\mathbf{r}_i - \mathbf{r}_j \right|} \\
    =&\sum_{i=1}^N\hat{H}_0^i + \hat{H}_2 = \hat{H}_0 + \hat{H}_2,
  \end{aligned}
\end{align}
where, in the first line, the first term is the kinetic energy and the second term is the potential energy due to nuclei. Both are summed over the contributions of individual electrons. The third term is the electron-electron interaction energy. We can divide the Hamiltonian into the non-interacting part, $H_0$, that is a sum over single-particle Hamiltonians $H_0^i$, and the interacting part, $H_2$, corresponding the Coulombic electron-electron interactions.

The single-particle states of the NV$^-$-center possess the same symmetries as $\hat{H}_0^i$. They are traditionally used to construct the many-body states. In practice they can be found with Kohn-Sham DFT, but a general idea can be obtained from analytic considerations. Working with the molecular orbital method \cite{coulson1957}, a group theoretical symmetry analysis (see Appendix) gives four single-particle orbitals built from the $4$ tetrahedratically symmetric dangling bonds of the defect: $a_N$, $a_1$, $e_x$, and $e_y$. $a_N$ and $a_1$ are fully symmetric and lower in energy, while $e_x$, and $e_y$ are degenerate and correspond to the same $2$-dimensional symmetry representation. The single-particle orbital energetics is presented in Fig. \ref{figure: sketch of the multiplet structure}.

\begin{figure}[!h]
  \includegraphics[scale=.45]{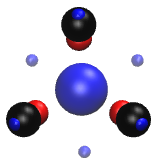}
  \includegraphics[scale=.45]{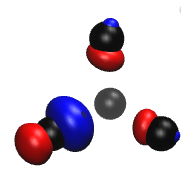}
  \includegraphics[scale=.45]{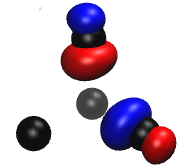}
  \caption{\label{figure: orbital depictions}The single particle orbitals $a_1$ (left), $e_x$ (middle) , and $e_y$ (right), shown from the direction of the high-symmetry axis of the NV$^-$-center. The black spheres represent atoms nearest to the vacancy; the atom in the middle is N, the others are C, and N is behind the C atoms in the figure. The negative (blue) and positive (red) parts of the orbitals are plotted. The orbitals are obtained from a spin unpolarized DFT simulation with a PBE correlation functional.}
\end{figure}

The shape of these orbitals from DFT simulations is shown in Fig. \ref{figure: orbital depictions}, and their mathematical formulation in a dangling bond basis is shown in Eq. (\ref{equation: sp states}) in Appendix. Density functional theory reveals that $a_1$, $e_x$, and $e_y$ lie within the band gap, while $a_N$ is below the valence band maximum and assumed to always be fully occupied. The negatively charged NV$^{-}$-center has thus $4$ electrons occupying the states in the band gap, and the open shell of the NV$^-$-center has the same single-particle energetics as diradical molecules \cite{rao2022}. 

To represent the many-body states, i.e. the eigenstates of $\hat{H}$, of the NV$^-$-center, we use the Slater-Jastrow wave functions:
\begin{equation}
  \label{equation:slater-jastrow wf, qualitative}
  \Psi_{SJ} =
  e^{J}
  \sum_n c_n
  D^\uparrow_n
  D^\downarrow_n
\end{equation}

The Jastrow factor $J$ \cite{drummond2004} includes one-, two-, and three-body functions of interparticle distances and is optimized to capture the electronic correlations of the system. The Slater determinants consist of the occupied electronic orbitals. For non-spin-dependent Hamiltonians we can represent the  Slater determinants as products of spin-up and -down determinants $D^\uparrow$ and $D^\downarrow$. The sum over Slater determinants is determined by the symmetries of the different many-body states.

Group theoretic symmetry analysis (see Appendix) can be used to find the many-body wavefunctions of different multiplet states. We restrict our consideration to different occupations of the single-particle orbitals within the band gap by $4$ electrons. The available wavefunctions for the states $3A_2$, $1E$, $1A_1$, and $3E$ can be found from Table \ref{table: many-body wave functions}. 
\begin{table}[ht]
  \caption{\label{table: many-body wave functions}%
  The many-body wavefunctions of the NV$^-$-center obtained with the symmetry projection method (see Appendix). The Jastrow factor is omitted for clarity.  First column shows the state, second the state's projection of the total spin along the symmetry axis of the NV$^-$-center, and third the wave function in terms of a sum of Slater determinants. Only the occupation of the single-particle orbitals within the band gap is shown in the Slater determinants. All the orbitals below the band gap are completely filled. The orbital denoted by $a$ below refers to $a_1$.}
  \begin{ruledtabular}
    \begin{tabular}{lcr}
      State & $S_z$ & Wave function\\
      \colrule
      $\Psi^{3A_2}_1$     & $1$    & \tiny{$D^\uparrow(ae_xe_y)D^\downarrow(a)$}\\
      $\Psi^{3A_2}_0$     & $0$    & \tiny{$\frac{1}{\sqrt{2}}\left(D^\uparrow(ae_x)
                         D^\downarrow(ae_y)- D^\uparrow(ae_y)D^\downarrow(ae_x)\right)$}\\
      $\Psi^{3A_2}_{-1}$  & $-1$   & \tiny{$D^\uparrow(a)D^\downarrow(ae_xe_y)$ }\\
      $\Psi^{1E}_x$       & $0$    & \tiny{$\frac{1}{\sqrt{2}}\left(D^\uparrow(ae_x)
                         D^\downarrow(ae_y) +  D^\uparrow(ae_y)D^\downarrow(ae_x)\right)$}\\
      $\Psi^{1E}_y$       & $0$    & \tiny{$\frac{1}{\sqrt{2}}\left(D^\uparrow(ae_x)
                     D^\downarrow(ae_x)- D^\uparrow(ae_y)D^\downarrow(ae_y)\right)$} \\
      $\Psi^{1A_1}$       & $0$    & \tiny{$\frac{1}{\sqrt{2}}\left(D^\uparrow(ae_x)
                         D^\downarrow(ae_x)+D^\uparrow(ae_y)D^\downarrow(ae_y)\right)$} \\
      $\Psi^{3E}_{x,1}$   & $1$    & \tiny{$D^\uparrow(ae_xe_y)D^\downarrow(e_x)$} \\
      $\Psi^{3E}_{y,1}$   & $1$    & \tiny{$D^\uparrow(ae_xe_y)D^\downarrow(e_y)$} \\
      $\Psi^{3E}_{x,0}$   & $0$    & \tiny{$\frac{1}{\sqrt{2}}\left(D^\uparrow(e_xe_y)
                         D^\downarrow(ae_x)-D^\uparrow(ae_x)D^\downarrow(e_xe_y) \right)$} \\
      $\Psi^{3E}_{y,0}$   & $0$    & \tiny{$\frac{1}{\sqrt{2}}\left(D^\uparrow(e_xe_y)
                         D^\downarrow(ae_y)- D^\uparrow(ae_y)D^\downarrow(e_xe_y)\right)$} \\
      $\Psi^{3E}_{x,-1}$  & $-1$   & \tiny{$D^\uparrow(e_x)D^\downarrow(ae_xe_y)$} \\
      $\Psi^{3E}_{y,-1}$  & $-1$   & \tiny{$D^\uparrow(e_y)D^\downarrow(ae_xe_y)$} \\
    \end{tabular}
  \end{ruledtabular}
\end{table}

In addition to the states shown in Fig. \ref{figure: sketch of the multiplet structure}, there is a doubly degenerate state $1E'$ and a fully symmetric $1A_1'$ state, consisting of determinants with single and double excitations, above the $3E$ triplet. These states are not included in Table \ref{table: many-body wave functions}, and are not in general relevant for the qubit operations of the NV$^-$-center. However, the coupling of the $1E$ and $1A_1$ singlets to singly and doubly excited determinants affect to the singlet energies \cite{jin2022}.

Quantum defect embedding theory predicts the weight of the determinants in the $1E'$ and $1A_1'$ states to account for approximately $3$\% of the wavefunctions \cite{bockstedte2018,sheng2022}, while  time-dependent DFT with hybrid DDH functional gives a decrease of $200-300$ meV to the energies when contributions of determinants $D^\uparrow(e_xe_y)D^\downarrow(e_xe_y)$, $D^\uparrow(e_xe_y)D^\downarrow(ae_x)$, and $D^\uparrow(e_xe_y)D^\downarrow(ae_y)$ are added to $1A_1$, $1E_y$, and $1E_x$, respectively \cite{jin2022}. While we hope the Jastrow factor (and the backflow function, see below) and the DMC method to account for the correlations in the simulations, we also test the coupling of the $1E$ and $1A_1$ states to the excited determinants. 

\subsection{Excited state energies}

We evaluate the vertical excitation energies $\Delta$ in the $3A_2$ ground state structure as
\begin{equation}
  \label{equation: excitation energy formula}
  \Delta_{state}=E_{state}-E_{3A_2},
\end{equation}
i.e. the vertical excitation energy of the state $E_{state}$ – with 'state' being either $1E$, $1A_1$, or $3E$ – is evaluated as the difference of the total QMC energies of the excited state and the $3A_2$ ground state in a given simulation cell.

The evaluation of the energies only in the $3A_2$ structure is a practical choice. Only the relaxation of structures at spin-up $3A_2$ and $3E$ states is possible with DFT, and the Stokes and anti-Stokes shifts for the excitation processes are small compared to available statistical accuracy in QMC. The absorption energy $\Delta_{3E}$ is known exactly from experiments. The differences in the structures of the ground state and the singlet states is assumed to be small due to similar electronic occupation in the single-particle picture. This assumption is supported by results in Ref. \cite{jin2022}, where time-dependent DFT with hybrid functionals allowed relaxation of the singlet states.  The structure of $1A_1$ was found to be very similar to the ground-state structure. The phonon sideband of $73$ meV in $1E\rightarrow 1A_1$ transition was found in the same study, which is only slighty larger than the available statistical accuracy in this QMC study.

The total energies of the wave functions in Table \ref{table: many-body wave functions} in a given simulation cell are computed with DMC, using a Jastrow factor (and backflow function, see below) optimized with VMC. The optimization of the Jastrow factor does not in principle improve the DMC results, but it makes the DMC sampling process more stable and reduces Monte Carlo errors, finite time step errors, finite population bias, and pseudopotential locality errors. 

\subsection{Fermion sign error, nodal surface and the backflow function}

The fermion sign problem limits the accuracy of the calculations. The Slater determinants constructed from the DFT single-particle orbitals fix the nodal surface of the wavefunction. As a practical solution to the sign problem the DMC simulation is constrained within nodal pockets having positive sign of the wave function. This fixed-node approximation limits the accuracy of the calculations to the accuracy of the nodal surface. We expect some cancellation of fixed node errors when evaluating the vertical excitation energies, but the many-body states of different symmetries can possess qualitatively different nodal surfaces, and hence we must explicitly study how to optimize the nodal surface for each state.

The choice of the DFT functional can alter the nodal surface. In bulk silicon, it has been found that hybrid PBE0 gives lower DMC total energies than the PBE functional or the Hartree-Fock method \cite{annaberdiyev2021}. Another way of tuning the nodal surface is to alter the restrictions for the electron densities in DFT simulations. For example, the orbitals can be solved self-consistently in either a spin-unrestricted (S.U.) or restricted (S.R.) calculation, where the former treats orbitals in different spin channels separately while the latter forces the orbitals and their occupations to be the same for both spins. S.U. calculation is needed to describe the $3A_2$ and $3E$ states with DFT, and in case it is used to prepare the trial wave functions for QMC, $D^\uparrow$ and $D^\downarrow$ in Eq. (\ref{equation:slater-jastrow wf, qualitative}) are constructed from spin-up and -down Kohn-Sham orbitals, respectively. However, the S.U. orbitals can lead to spin contamination \cite{baker1993}, and hence it can be that S.R. orbitals are needed at least for the singlets $1E$ and $1A_1$ and the zero spin projections $S_z=0$ of the triplet states, where the number of up- and down-spin electrons is the same. We can also solve the orbitals in the $3E$ excited state occupation with constrained occupation DFT. However, we found that the degeneracy of the $e_x$ and $e_y$ orbitals was broken when forcing excited state orbital occupation. This means that the symmetries of the $e$ orbitals required to construct the many-body states appropriately was broken, and hence we did not use this method.

Also optimization of the coupling coefficients of the $1E$ and $1A_1$ states to the singly and doubly occupied determinants affects the nodal surface. As the states only couple to determinants of equal symmetry, the nodal topology should remain the same, therefore not altering the variational principle (see below). 

The final way that the nodal surface was optimized in this study was to include a backflow correction \cite{kwon1993,kwon1998,lopez2006} to the many-body wavefunctions. The backflow includes parametrized shifts to the particle coordinates as $\mathbf{r_i}\rightarrow \mathbf{r}_i +\xi_{i}(\mathbf{R})$, so that the coordinate of particle $i$ is dependent on the coordinates of the rest of the particles. In this study, we use backflow functions with one-, two-, and three-body terms. The inclusion of a backflow function to the wave function increases the variational freedom of the wave function and optimization of the backflow improves the accuracy of the wave function and the nodal surface.

The variational principle is not guaranteed to hold in DMC for the excitations, even in cases where the excitation has a symmetry orthogonal to the ground state symmetry. Foulkes \textit{et al.} \cite{foulkes1999} have shown that excited states transforming as multidimensional irreducible representations of the symmetry group of the Hamiltonian do not in general fulfill the variational principle. Thus the $1A_1$ has a variational principle, but $1E$ and $3E$ do not. However, under special circumstances a weaker or even the original variational principle can remain for multidimensional excitations \cite{foulkes1999}. In the Appendix, we show that the variational principle applies for the $y$-components of the spin-up $3E$ and $1E$ states depicted in Table \ref{table: many-body wave functions}. Thus we can use for these wavefunctions all the methods described above to optimize the nodal surfaces by minimizing energy. 

It should be noted, however, that with exact nodal surface the DMC energy should be exact also for the excited energies, whether the variational principle applies or not. The fixed-node errors are quadratic in the error in the nodal surface when the variational principle applies, and linear when it does not \cite{hunt2018}. Therefore, with a nodal surface that is close to correct, we should be able to get approximate excited energies even when variational principle does not hold \cite{needs2020}.

\subsection{Finite size effects}\label{subsec: fs effects}

There are a number of finite-size errors present in the calculation of total energies of the simulation cells of the NV$^{-}$-center. Particle momentum is evaluated in a discrete grid instead of continuum. There are errors in simulations of the charged supercell of the NV$^{-}$-center, where the extra negative charge has to be compensated with a positive jellium background. Furthermore, the simulation of a defect in a periodic cell introduces errors originating from the defect interacting with its periodic images.

These errors should largely cancel when we evaluate vertical excitation energies. Results from DFT, and some more advanced methods, can shed light on the behaviour of the listed finite size effects. In the DFT study of Ref. \cite{gali2008}, a $511$-atom simulation cell was found to have convergent charge density and excitation energies. The structural changes in the system were negligible when moving from a $215$-atom to a $511$-atom simulation cell. In a very recent study with time-dependent hybrid DFT, $215$- and $511$-atom simulation cells gave vertical excitations within $20$ meV \cite{jin2022}. A study by Ma \textit{et al}. found the relevant excitation energies of the $255$-atom FCC simulation cell to be convergent with respect to cell size \cite{ma2010}. Our hybrid DFT results for the absorption energy, zero-phonon line, Stokes and anti-Stokes shifts for the $3A_2$ and $3E$ states in Fig. \ref{figure: dft structure results} show that already a $215$-atom simulation cell gives convergent results, with differences of at most $15$ meV between the $215$- and $511$-atom simulation cells. These results hint that errors due to momentum discretization, periodic point charge of the defect, defect relaxation and defect-defect interactions should be small with the $215$-atom simulation cell.  

The Coulomb interaction cannot be used for finite simulation cells to represent infinite systems, as the sums over particle images do not converge absolutely. Instead, model interactions mimicing the Coulomb interaction in an infinite lattice as closely as possible must be used. The use of Ewald interaction in particular can cause the exchange-correlation holes to interact with their periodic images \cite{kent1999}, also in excitations \cite{hunt2018}. Our results with both model periodic coulomb (MPC, used both in particle branching and energy evaluation in DMC) and Ewald interactions give the same excitation energies within QMC errorbars, hence verifying that at least the interactions of the exchange-correlation holes with their periodic images should be largely canceled \cite{kent1999}.

The remaining finite size effects that need to be considered are the electrostatic finite-size errors arising from the difference in the quadrupole moment between ground and excited electronic states, and the finite-size errors originating from the relaxation of the atomic positions in the simulation cell. Both scale as $L^{-3}$, where $L$ is the linear size of the supercell, so that with particle number $N$, $L^3 \sim N$. There are also nonsystematic, quasirandom finite size effects; oscillations in the electron pair density and the momentum grid are forced to be commensurate with the simulation cell, leading to random bias in energy.

We estimate infinite system size energies $E(\infty)$ by fitting energies $E(N)$ from simulation cells with particle number $N$ as
\begin{equation}
  \label{equation: extrapolation formula to infinite system size energies}
  E(N)=E(\infty)+\frac{A}{L^3},
\end{equation}
This simple form of the fitting function should avoid overfitting of the results. The exponent of the finite-size error function is chosen based on the above considerations.

Finite size effects should be studied separately for the backflow-corrected wavefunctions as well. Usually the finite-size errors with backflow are slightly larger than without the backflow. However, the change in finite size effects due to backflow are assumed to be small, and the long-range finite size effects expected to be largely canceled in excitation energies \cite{holtzmann2016}. Because of these arguments, and also due to computational limitations, we only estimate the effect of the backflow in the $63$-atom simulation cell, and extrapolate the QMC results to infinite system size based on Slater-Jastrow results in $63$- and $215$-atom cells. 

\section{Computational details}

\subsection{Simulation cells}

The simulations were done in negatively charged cubic $63$- and $215$-atom ($254$- and $862$-electron) simulation cells to get the main set of results. 

\subsection{Structural relaxation}

The structures were relaxed with a HSE06 correlation functional \cite{heyd2003} and PAW pseudopotentials \cite{kresse1999} by using the Vienna ab initio simulation package (VASP) \cite{kresse1996a,kresse1996b}. In a spin-unrestricted calculation the electrons occupied orbitals so that the self-consistent charge density corresponded to that of the spin-up $3A_2$ state. This enabled us to find the ground-state structure. For the $63$-atom simulation cell we used $2\times 2\times 2$ $\mathbf{k}$- mesh, but in the $215$-atom simulation cell the use of only the $\Gamma$-point gave convergent structures with respect to the sampling of the $\mathbf{k}$-mesh. The plane-wave cutoff of the electronic orbitals was $500$ eV, and the structures were relaxed with a $0.002$ eV/Å force convergence criterion.

\subsection{Orbital generation}

The orbitals were generated with Quantum Espresso's PWSCF package \cite{giannozzi2020} by running a self-consistent DFT calculation for the relaxed simulation cells. Norm-conserving pseudopotentials were used. This combination of PWSCF and pseudopotentials has a user-friendly interface with the CASINO QMC program which was the reason for using the package. We used both PBE and HSE06 correlation functionals. The plane-wave cutoff was tested separately here for $1$ meV/atom convergence, and a cutoff of $2700$ eV was used. As mentioned above, both S.U. and S.R. calculations were performed to later test the quality of the trial QMC wave functions composed of the self-consistent single-particle orbitals. 

The orbitals were always generated with only a single $\mathbf{k}$-vector at $\Gamma$. The finite simulation cell includes artificial dispersion to the single-particle orbitals, possibly affecting the vertical excitation energies evaluated with DFT and QMC. However, we tested the effect of dispersion by evaluating multiple trial wave functions with randomly shifted $\mathbf{k}$ vectors in the $63$-atom simulation cell, calculated the twist averaged $\Delta_{3E}$ with QMC, and found the energy not to differ within statistical accuracy of $50$ meV from the energy evaluated only with the $\Gamma$ wavefunction.   

\begin{figure}[!h]\hspace*{-.41cm}
  \includegraphics[scale=.49]{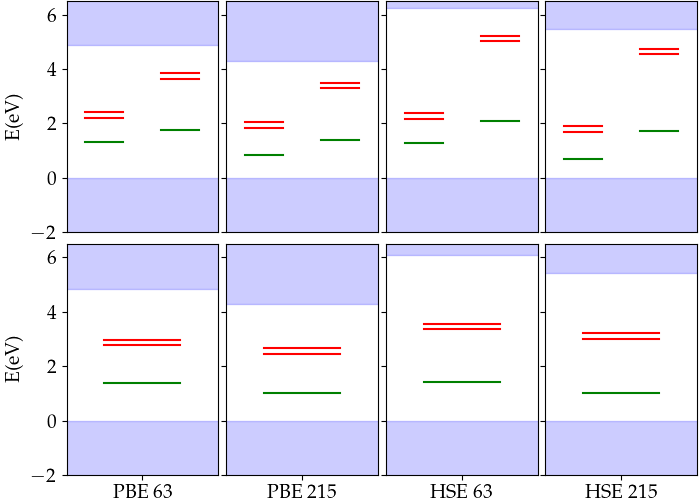}
  \caption{\label{figure: orbital bands} Energies of the S.U. (top row) and S.R. (bottom) single particle orbitals in $63$- and $215$-atom simulation cells, computed with PBE and HSE06 correlation functionals. The green line represents the energy of the orbital with $a$-symmetry, and red the energies of orbitals with $e$-symmetry. The red lines representing degenerate states are vertically displaced for presentative purposes.}
\end{figure}

Figure \ref{figure: orbital bands} presents the energies of the single-particle orbitals in the band gap of diamond obtained from $8$ different DFT simulations; the energies were obtained with $2$ different simulation cell sizes, $2$ different correlation functionals, and by either allowing the orbitals of different spin channels to vary or forcing them to be the same. The final results for the QMC excitation energies should be evaluated with both cell sizes, but we should choose whether we use PBE or HSE, and S.U. or S.R. orbitals to construct the many-body wavefunctions.

\subsection{QMC}

The QMC simulations were done for the ground and excited state wavefunctions with the CASINO package \cite{needs2020}, by first running an optimization of the wave function in VMC level. The optimizations were done with $40\ 000$ electronic VMC-sampled configurations with an energy minimization algorithm in all simulation cell sizes. For the Jastrow factor this provided convergence of the wave function energy with a precision of $1-2$ meV/atom in both cell sizes for the Jastrow factor. Tests with $320\ 000$ configurations did not improve the optimized energy. We tested to optimize $3A_2$ Slater-Jastrow-backflow wave function also with $120000$ configurations, and found no gain in energy within the statistical errorbars of $150$ meV for the whole simulation cell. The coupling of the $1E$ and $1A_1$ states to the singly and doubly excited determinants were optimized together with the Jastrow factor with $80000$ configurations, without any gain in energy by adding the number of configurations to $10$-fold.

\begin{figure}
  \includegraphics[scale=.67]{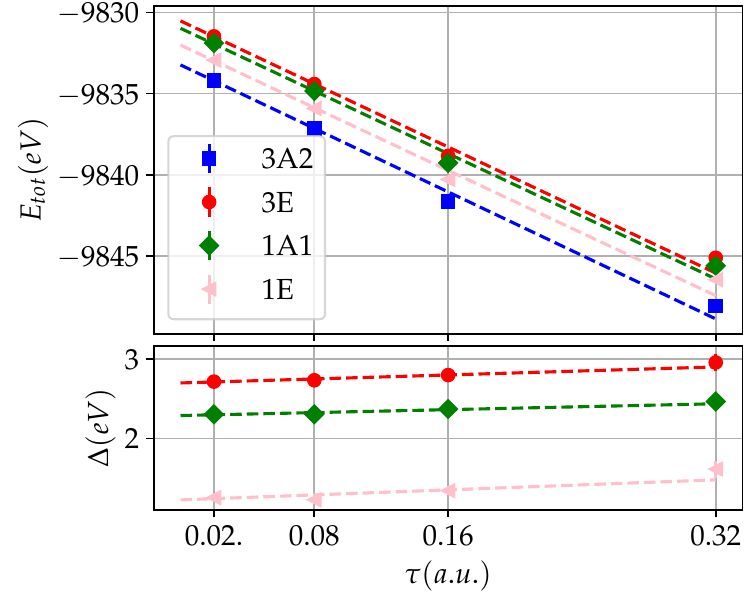}
  \caption{\label{figure: dtdmc} Total energies (top) and vertical excitation energies (bottom) of the Slater-Jastrow many-body states of the NV$^{-}$-center in a $64$-atom simulation cell as a function of the DMC time step, with DMC error bars. The dashed lines represent linear fits to the state energies.}
\end{figure}

The optimized wave functions were then taken to DMC simulations, where final results were obtained by running DMC with two different time steps and extrapolating to zero time step. Previous excited-state QMC calculations found significant cancellation of finite time-step errors \cite{hunt2018} in bulk materials, and we found similar features. Figure \ref{figure: dtdmc} shows the total energies of the many-body states and vertical excitation energies in a $63$-atom simulation cell as a function of the DMC time step. We can see that while the dependence of the total energy on the time step is large, with $>10$ eV difference between time steps of $0.32$ and $0.02$ a.u., the dependence of $\Delta$ is much smaller, with corresponding difference of less than $400$ meV. We chose to use time steps of $0.08$ and $0.02$ a.u. for extrapolations to zero time step. We could have used larger time steps to save computation time, but we found that by doing so, finite population bias and even population explosions occurred in $215$-atom simulation cells, making the accumulation of reliable results more difficult. With the smaller time steps the DMC was run with twice as many imaginary time steps, and with four times more walkers than with the larger time step. This ensures the optimal extrapolation that also minimizes the finite population bias \cite{lee2011}. 

The DMC equilibration period with a chosen time step was chosen so that the root mean square distance of the diffusion of particles during equilibration was $3-4$ times greater than the Wigner-Seitz radius of the simulation cell, which should suffice according to \cite{needs2020}. 

We also tested different pseudopotentials to verify the validity of the core treatment in the simulations. We used Dirack-Fock (DF) pseudopotentials \cite{trail2005a,trail2005b} and energy consistent correlated electron pseudopotentials (eCEPPs) \cite{trail2017} by Trail \& Needs. Both pseudopotentials leave $4$($5$) valence electrons for individual C (N) atoms in the simulation.

We found that the results for $\Delta$ in $63$-atom simulation cell with both pseudopotentials were equivalent within the standard deviations of the results. This justifies the use of either one of the pseudopotentials. We chose to use DF pseudopotentials, as they required smaller plane-wave cutoffs for DFT orbitals and demonstrated smaller DMC time step errors. The calculations were performed with the T-move scheme by Casula \cite{casula2006} to establish the variational principle for the DMC energies with pseudopotentials. 

Model periodic Coulomb (MPC) interactions \cite{kent1999} were used to model the Coulomb interactions in the periodic simulation cell. As mentioned above, we found no difference in vertical excitation energies in $64$-atom simulation cell between calculations done with Ewald and MPC interactions, but chose MPC due to computational speedup it provides.

\section{Results}\label{section: results}

\subsection{Structural relaxation}

\begin{figure}[!ht]
  \includegraphics[scale=.5]{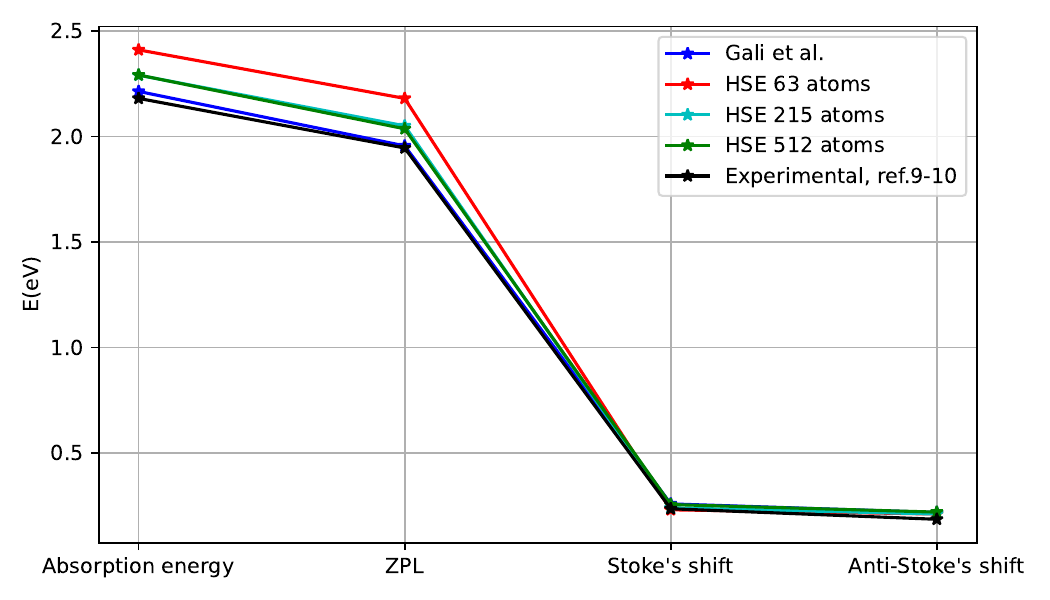}
  \caption{\label{figure: dft structure results}DFT results of the absorption energy, zero-phonon line, Stokes and anti-Stokes shifts obtained with HSE06 hybrid functional for cubic $63$-(red), $215$-(cyan), and $512$-(green) atom simulation cells. For comparison, also HSE06 results with $512$-atom simulation cell by Gali \textit{et al.} \cite{gali2009} and experimental results are presented.}
\end{figure}

Figure \ref{figure: dft structure results} shows our DFT results from calculations with VASP for the absorption energy, zero-phonon line, Stokes and anti-Stokes shifts for the $3A_2$- and $3E$-states in $63$-, $215$-, and $511$-atom simulation cells with HSE06 correlation functional. The results were obtained with a single $\mathbf{k}$-point and a plane-wave cutoff of $500$ eV.  Also results from another work with HSE06 and $512$-atom simulation cell are presented \cite{gali2009}, as well as experimental reference results \cite{davies1976}. We can see that our HSE06 results converge already with $215$-atom simulation cell. Our $511$-atom results are approximately $80$ meV higher than results by Gali \textit{et al}. \cite{gali2009} for absorption energy and zero-phonon line, but the Stokes and anti-Stokes shifts agree well. We take the results accurate enough against experiment to validate the use of the ground-state structure in further simulations.

\subsection{Generation of single-particle orbitals}

With the relaxed ground-state structures, we solved the single-particle orbitals with norm-conserving DF pseudopotentials and the PWSCF package. The results of the single-particle energies of the localized orbitals are already presented in Fig. \ref{figure: orbital bands}. We got the same $\Delta_{3E}$ with the new pseudopotentials and the HSE06 functional as in the previous phase of the calculations, where we used PAW pseudopotentials. Fig. \ref{figure: orbital bands} shows the well-known result of PBE underestimating the diamond band gap, while HSE06 with the $215$-atom simulation cell gives a band gap very close to the experimental value of $5.47$ eV.  

\subsection{QMC results}

\subsubsection{Total energies}

\begin{figure*}\hspace*{-1cm}
  \includegraphics[scale=.6]{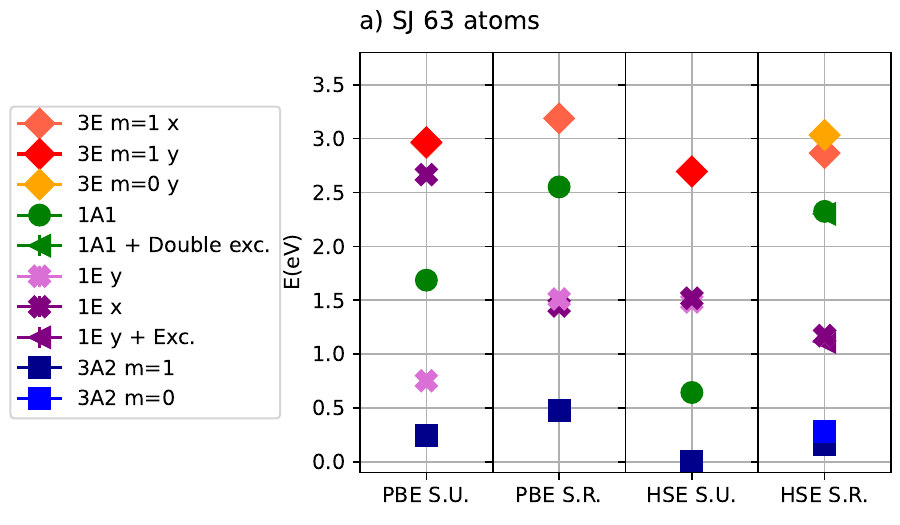}
  \includegraphics[scale=.6]{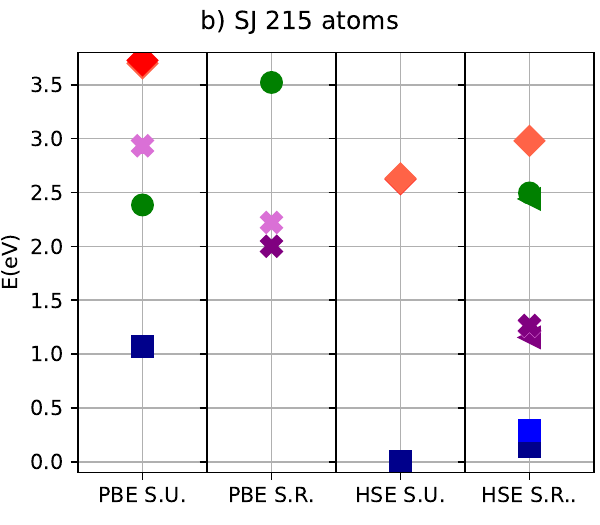}
  \includegraphics[scale=.6]{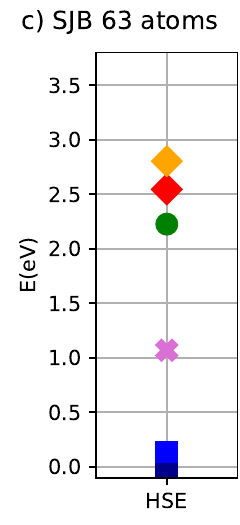}
  \caption{\label{figure: 64 atom multiplet results}Results of the QMC energies of the states of the NV$^{-}$-center against the lowest obtained ground state energy in the $63$- (a) and 215-atom (b) simulation cells with the Slater-Jastrow (SJ) wave function, and in the $63$-atom cell with a backflow correction (c). The SJ wave function results are computed with electronic orbitals prepared using PBE and HSE06 correlation functionals in S.U. and S.R. DFT calculations. The results with backflow are computed with HSE06 S.R orbitals for the $S_z=0$ and with HSE06 S.U. orbitals for the $S_z=1$ spin manifold. Results for the states $3A_2$ (blue), $1E$ $x$-and $y$-components (violet and pink), $1A_1$ (green), and $3E$ $x$-and $y$-components (orange and red) are shown. For the triplets represented with HSE06 S.R. orbitals, also the triplet states with $S_z=0$ spin projections are shown (lighter orange). In the $216$-atom simulation cell, we do not present results for the triplets with PBE S.R. orbitals and for the singlets with HSE06 S.U. orbitals. For some of the results, such as for the energies of the $x$- and $y$-components of the $3E$ state with HSE06 S.U. orbitals, the energies are overlapping although it is not visible from the figure. }
\end{figure*}

Figure \ref{figure: 64 atom multiplet results} shows DMC results of the vertical excitation energies $\Delta$ of states in $3A_2$, $1E$, $1A_1$, and $3E$ for a representative set of wavefunctions. The Slater-Jastrow results are for $63$- (a) and $215$-atom (b) simulation cells. Results with backflow in the $63$-atom cell are also presented (c). In both cell sizes, excitations are calculated with $4$ different sets of DFT orbitals, as distinguished in Fig. \ref{figure: orbital bands}: PBE S.U., PBE S.R., HSE S.U., and HSE S.R., presented in this order in the panels (a) and (b). The $S_z=0$ projections of the triplet states are shown only with HSE06 S.R. orbitals. In panel (c), the energies of the $S_z=0$ states are computed  with S.R. orbitals and the $S_z=1$ states with S.U. orbitals. In each panel the results are shown relative to the lowest energy obtained for the $3A_2$ state in the given cell.

We can see that the lowest $3A_2$ ground-state energies are obtained for the $S_z=1$ projection with S.U. HSE06 orbitals in both cell sizes. According to the variational principle these sets of orbitals have the lowest fixed-node errors. We tried to optimize the nodal surface errors further by altering the exchange mixing parameter $\alpha$ in the HSE06 functional, but found no changes in $3A_2\rightarrow 3E$ excitation with exchange parameters between $\alpha=0.15$-$0.25$ within the statistical errorbar of $0.7$ eV.

We estimated the DMC energies of the $S_z=0$-projection of the $3A_2$ state with HSE06 S.R. orbitals. This gave $\sim 230$ meV higher energies with SJ wave functions than for the $S_z=1$ projection of the $3A_2$ state with HSE06 S.U. orbitals. Also the energy of the $S_z = 1$-projection of the $3A_2$ state with S.R. orbitals gave higher energies than with S.U. orbitals in the SJ level. From these results we know that the S.R. orbitals cause larger nodal surface errors than S.U. orbitals. For the $S_z=0$ projection, also the determinants are different, which may increase the nodal errors. The $S_z=0$ projection of the ground state may also suffer from spin contamination with the singlet states in the $S_z = 0$ manifold \cite{huang1998}. If we used S.U. orbitals for the $S_z = 0$ projection of the ground state, the estimated energies were more than $2$ eV higher than the lowest ground state energy, indicating stronger spin contamination effects.

When using the backflow, the energy of the $S_z=1$ projection of the ground state with S.R. orbitals exactly overlaps with the corresponding energy with S.U. orbitals, as seen in Fig. \ref{figure: 64 atom multiplet results} (c). Hence backflow corrects the possible errors due to nodal surfaces or spin contamination. The possible reduction in spin contamination would align with the findings of Huang \textit{et al.} \cite{huang1998} that more correlated wave functions reduce the spin contamination.  

The different orbitals and spin projections of the $3E$ state provided similar energy discrepancies as with the $3A_2$ state, and the backflow was seen to reduce the discrepancies also in this case. However, the effect of backflow was slightly smaller for the $3E$ state. 

The orbitals with lowest fixed-node errors can also be chosen for the excitations that maintain the variational principle (see Appendix). For the $y$-component and $S_z=1$-projection of the triplet $3E$-state, we see, as noted above, that again the S.U. HSE06 orbitals give the lowest energy in both cell sizes. As anticipated, we encountered problems when simulating the singlet states with S.U. orbitals: in the $63$-atom cell, the ordering of the singlets is qualitatively wrong with S.U. HSE06 orbitals. Also the degeneracy of the $1E$-state with S.U. PBE orbitals is broken with a split between $x$- and $y$-components of $\sim 2$ eV. The S.R. orbitals with both functionals maintain the $1E$ degeneracy in both cell sizes, and also the ordering of the singlets matches the well-known picture with S.R. orbitals. Thus we restrict the description of the singlet states to S.R. orbitals. Within this restriction we find that HSE06 orbitals give lower singlet energies in both cells. 

The conclusion from the results of Fig. \ref{figure: 64 atom multiplet results} is thus that one should use HSE06 orbitals to construct the Slater determinants for all of the many-body states, S.U. orbitals for the $S_z=1$ projections of the triplet states and S.R. orbitals for the multideterminant states with $S_z=0$. 

Finally, Fig. \ref{figure: 64 atom multiplet results} shows DMC results for the $\Delta_{1E}$ and $\Delta_{1A_1}$, when the coupling of the states to singly and doubly excited determinants, respectively, are added and optimized for the wavefunctions. In both cell sizes, we can see that the coupling is not significant. This is understandable, as DMC projects out the lowest state of a certain symmetry, and the coupling to determinants of equivalent symmetry does not alter the symmetry of the states.

\subsubsection{Infinite system size energies}

Figure \ref{figure: results, extrapolations and backflow} shows the ground state (a) and vertical excitation (b) energies in both cell sizes, with extrapolation to infinite system size according to Eq. (\ref{equation: extrapolation formula to infinite system size energies}) and with the backflow corrections. Based on the findings of the previous section, the triplet-to-triplet excitation is estimated with $S_z=1$ projections of the wave functions represented with S.U. orbitals, but the triplet-to-singlet excitations are computed with S.R. orbitals and by using the $S_z=0$ projection of the ground state, so that the Kohn-Sham ansatzes of the ground and excited states are coherent when estimating excitation energies. The backflow functions are included in the smaller $63$-atom simulation cell and the differences between results with and without the backflow are added to the extrapolated results. HSE06 was used to prepare the orbitals for all of the states. For $1E$ and $3E$ states we computed the energies of the $y$-components of the degenerate states.

%With respect to the different ground state representations, we can draw the same conclusion from Fig. \ref{figure: results, extrapolations and backflow} as from Fig. \ref{figure: 64 atom multiplet results}: with Slater-Jastrow wave functions, the S.R. orbitals provide larger ground state energies, and $S_z=0$ projection of the ground state gives larger energy than the $S_z=1$ projection, but the backflow correction diminishes the discrepancies.

%To inspect whether it makes a difference to calculate the vertical excitation energies of the singlet states with S.R. orbitals by using ground state wave function with $S_z=0$ or $S_z=1$ spin projections constructed from S.R. or S.U. orbitals, we have calculated the infinite size extrapolations and backflow corrections for the singlet excitation energies with both ground state representations in Fig. \ref{figure: results, extrapolations and backflow}.

Fig. \ref{figure: results, extrapolations and backflow} (a) shows that, while the representations of the ground state with S.R. orbitals remain higher in energy than with S.U. orbitals when SJ wave functions are used, backflow mainly corrects the discrepancies. The $S_z=0$ projection of the $3A_2$ state remains little higher in energy than the other two representations even after the backflow correction, but this discrepancy is very small with the given statistical accuracy of $\sim 100$ meV. The finite-size effects of the different ground state representations are identical. These results indicate that neither the spin projection of the $3A_2$ wave function nor the spin restrictions of the orbitals (with $S_z=1$ wave function projections) makes a difference when estimating the vertical excitation energies.

\begin{figure}[!htb]
  \hspace*{-.9cm}
  \includegraphics[scale=.55]{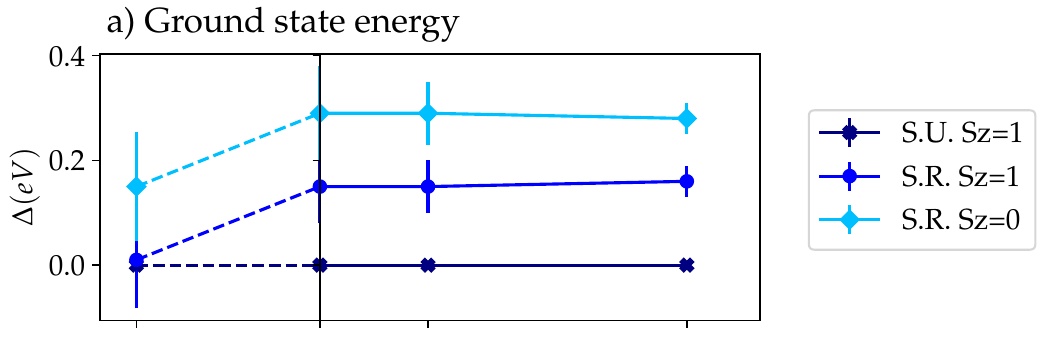}
  \hspace*{-.9cm}
  \includegraphics[scale=.55]{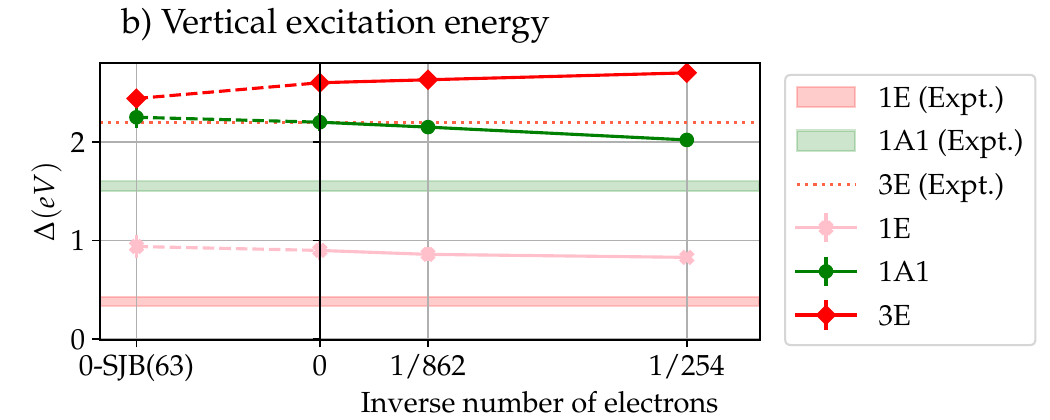}
  \caption{\label{figure: results, extrapolations and backflow} Energies compared against the $S_z=1$ projection of the $3A_2$ ground state with HSE06 S.R. orbitals. Results are presented for Slater-Jastrow wave functions as a function of the inverse number of electrons in the simulation cell (right panels), for SJ wave functions with energies extrapolated to infinite system size (vertical line at $0$), and for the extrapolated results with a backflow correction (left panel, results at $0-SJB(63)$). In (a), the energies of the $S_z=0$ and $Sz=1$ projections of the $3A_2$ state, represented with S.R. orbitals, are compared. In (b), the vertical excitation energies of the $1E$-, $1A_1$-, and $3E$-states, drawn with pink, green and red, respectively, are presented. The singlets were represented with S.R. orbitals, and the $3E$ with S.U. orbitals and the $S_z=1$ projection of the triplet. }
\end{figure}

In Fig. \ref{figure: results, extrapolations and backflow} (b) the $1E$ singlet has neglible finite-size effects within the statistical accuracy of the calculations. The $1A_1$ singlet has an excitation energy of $2.02(3)$ eV in the $63$-atom cell, while the energy extrapolated to infinite system size is $2.2(1)$ eV. The extrapolated energy of the $3E$ state is $100$ meV lower than the energy of the $63$-atom cell, which is barely visible in the current statistical accuracy of $\sim 70$ meV for the $3E$ state. An earlier study found a value of $70$ meV as the finite-size correction due to electrostatics for the triplet-to-triplet transition \cite{londero2018}. Overall the finite-size effects are small, with the $215$-atom simulation cell having the same energies as the extrapolated results within the statistical accuracy. The backflow corrections are also small; only for the vertical excitation energy of the $3E$ state we see a change in energy due to backflow ($160$ meV) larger than the statistical uncertainty.

After infinite-size extrapolation and backflow correction, the absorption energy corresponding to $3A_2 \rightarrow 3E$ from the QMC simulations overestimates the experimental value by $\sim 200$ meV, with an errorbar of $100$ meV. The vertical excitation energy results of the singlets are overestimating more when compared to experiment, but it should be noted that the singlet-to-singlet transition, known from experiments to be $1.19$ eV, is correct within the errorbars. It is possible that the cancellation of fixed-node errors is not as large in vertical excitation energies $\Delta_{1A_1}$ and $\Delta_{1E}$ as in $\Delta_{3E}$, as the singlets may have stronger multireference character than the ground state triplet, as found in \cite{chen2023} for the $1E$ state. This could explain the larger overestimation of the excitation for the singlets. On the other hand, we found the overestimation of the excitation energy of the $3E$ state comparable to the overestimation of the singlet energies when S.R. orbitals were used (results with HSE06 S.R. orbitals in Fig. \ref{figure: 64 atom multiplet results} (a) and (b)). Hence further optimization schemes for the singlet orbitals could be inspected to reduce the nodal surface errors of the singlet states.  

%The use of backflow reduces nodal surface errors and improves the vertical excitation energies. Interestingly, the $S_z=0$-projection of the $3A_2$ state has the same orbitals and determinants as the singlet states, but its energy is corrected by the backflow to correspond exactly to the $S_z=1$-projection, as seen in Fig. \ref{figure: 64 atom multiplet results}. This could mean that, as backflow is not as efficient for the singlet energies, accurate energies for the singlets would require larger multideterminant expansions.  

\subsubsection{Comparison to other work}

\begin{figure*}
  \includegraphics[scale=.6]{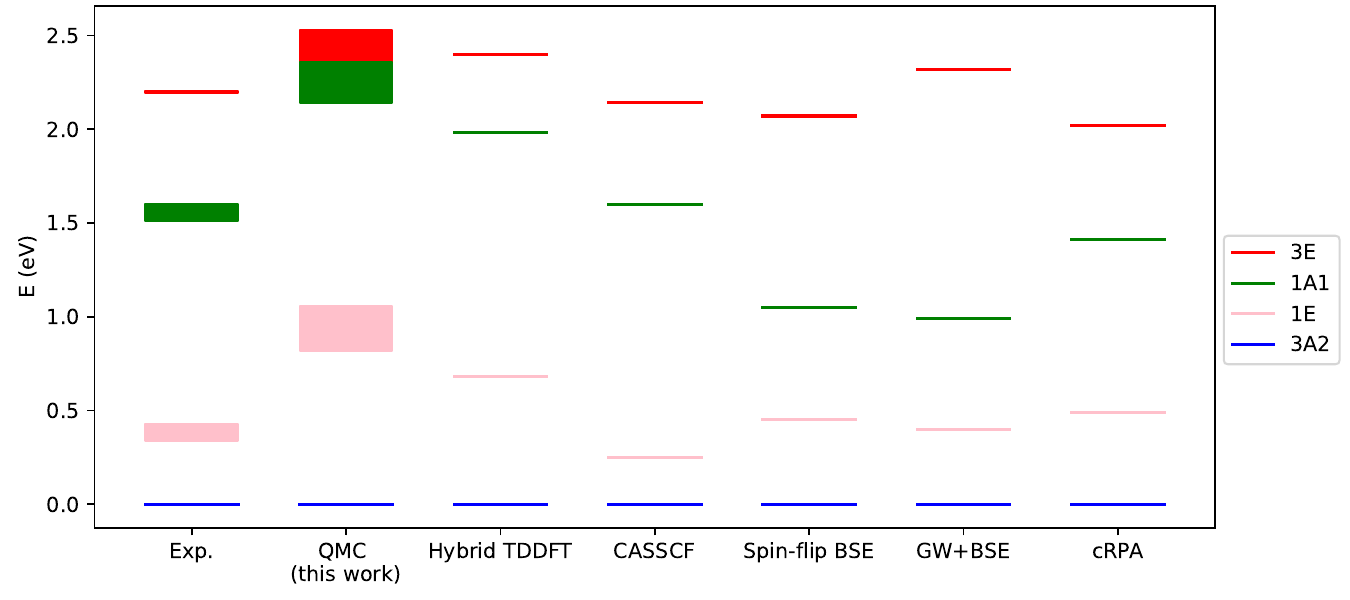}
  \caption{\label{figure: theoretical comparison} Vertical excitation energies of the $1E$ (pink), $1A_1$ (green), and $3E$ states (red) against the ground state (blue). Results are shown from experiments \cite{davies1976,rogers2008,goldman2015,goldman2015b} and QMC (this work), compared against results from a number of theoretical methods: hybrid time-dependent DFT (hybrid TDDFT) \cite{jin2022}, complete active space multiconfiguration SCF (CASSCF) \cite{bhandari2021}, spin-flip Bethe-Salpeter equation approach (Spin-flip BSE) \cite{barker2022}, GW approximation and BSE (GW+BSE) \cite{ma2010}, and constrained random-phase approximation (cRPA) \cite{bockstedte2018}.}
\end{figure*}

Figure \ref{figure: theoretical comparison} shows the DMC vertical excitation energies from Fig. \ref{figure: results, extrapolations and backflow}, after extrapolation to infinite system size and the backflow correction, against experimental values and results from other theoretical methods. Previous experimental studies have identified the zero-phonon-line transitions between the triplets \cite{davies1976} and singlets \cite{rogers2008} that, together with the energy interval between $344$ and $430$ meV for the $3E\rightarrow 1A_1$ transition from the study in publications \cite{goldman2015,goldman2015b}, allow one to draw the estimated vertical excitation energies for the states in Fig. \ref{figure: theoretical comparison}.

Hybrid time-dependent density functional theory (Hybrid TDDFT) was used to compute the vertical excitation energies in Ref. \cite{jin2022}, that also showed the structural relaxations of the singlet states to have only a minor effect on the energy, hence not explaining the current overestimation of the singlet energies by QMC. Hybrid TDDFT is fairly accurate, but shows overestimation in the vertical excitation energies. In Ref. \cite{bhandari2021}, hydrogen-passivated atomic clusters were used to model the NV$^-$center with complete active space self-consistent field (CASSCF) calculations, where multideterminant wave functions were optimized in a space of $6$ virtual orbitals populated by $6$ electrons. Two extra orbitals were found to localize to the defect region and to give important contributions in description of the correlation of the many-body states. The identification and inclusion of more orbitals to the multideterminant expansions in QMC trial wave functions could also very well improve the nodal surfaces of this study. Overall, CASSCF gives a good match to experimental energies, but underestimates the energy of the $1E$ state and therefore overestimates the singlet-to-singlet transition energy. Refs. \cite{barker2022} and \cite{ma2010} employ Bethe-Salpeter equation (BSE) approaches, with the former allowing also spin flips. These studies have a good match with experiment in the energy of the $3E$ state, and get similar singlet energies, underestimating the $1A$ state. Interestingly, the study employing spin-flip BSE found that the energy of the $3E$ state with S.R. orbitals was $300$ meV higher than with S.U. orbitals, which is similar to what we found at the SJ wave function level. Constrained random-phase approximation of \cite{bockstedte2018} works in a Kohn-Sham CI basis and uses a screened effective interaction potential, and has a good match against experiment. 

Fig. \ref{figure: theoretical comparison} shows that $\Delta_{3E}$ from QMC is in line with the other methods and experiment with minor variation, but that the vertical excitation energies of the singlets are larger than for any of the other methods and experiment. However, the singlet-to-singlet transition of $1.19$ eV from experiments is well repeated by QMC. Other methods able to capture both the triplet-to-triplet and singlet-to-singlet transitions with similar accuracy include only the hybrid time-dependent DFT.

The DMC directly simulates the many-body wave function, as opposed to methods utilizing effective Hamiltonians. The only sources of errors in DMC simulations are in the nodal surface. We are explicitly simulating the electron correlations in simulation cells up to $215$ atoms and extrapolating to infinite system size, which has not been done before by an explicitly correlated method. The variational principle that applies for the many-body states (see Appendix) also allows systematic improvement of the results with, e.g. larger multideterminant expansions in the trial wave functions for reducing the nodal  surface errors.

\section{Conclusions and Discussion}\label{section: discussion}

In this work we examined best practices and workflows for using QMC in the study of optical properties of point defect qubits. By using the NV$^-$-center in diamond as a test system, we calculated the vertical excitation energies of the many-body states relevant for qubit operation. The results were extrapolated to infinite system size and backflow corrections were included to the extrapolated excitation energies. The final results showed overestimation in vertical excitation energies when compared against experimental estimates, but the singlet-to-singlet transition energy from QMC, $1.2(1)$ eV, was correct against the experimental value of $1.19$ eV. The triplet-to-triplet absorption energy from QMC, $2.43(9)$ eV, was overestimated by approximately $0.23(9)$ eV against experiment.

We have removed the finite DMC time step errors, finite-size errors, and tested different pseudopotentials. We use backflow to optimize the nodal surfaces. The energies were computed by explicitly simulating the correlations of electrons in the defect and the surrounding medium. Remaining sources of error in the simulations should only be in the nodal surface, which makes our simulations very thorough as compared to many previous computational studies of the NV center in diamond. However, it was found that while the qualitative features were well reproduced and some transitions very accurately calculated, DMC was not able to match the triplet-singlet gaps with experimental values when only the orbitals in the band gap were used in the multideterminant expansions.  

DMC is devised for ground-state simulations, so the nodal surface errors could be larger for the excited states. This would explain the overestimation of the excitation energies in this study, even with the measures taken to optimise the nodal surfaces. The use of HSE06 hybrid correlation functional for obtaining the electronic orbitals was found to improve the wave functions when compared to PBE orbitals. Also S.U. orbitals were found to give better description of the triplet states with Sz=1 spin projection. 

%All of the excited state energies were overestimated against the ground state energy when compared against experiment, so improvements in the wave functions should be made. The variational principle enables systematic studies on how to improve the wavefunctions further. We have found that S.R. orbitals should be used for states with $S_z=0$, while S.U. orbitals provide better description of the states with $S_z=1$. Also the use of HSE06 electron orbitals improves the wave functions when compared to PBE orbitals.

The use of different set of orbitals for the singlets and the $S_z=0$ projections of the triplets than for the $S_z=1$ projections of the triplets can prove to be problematic for the generalization of the QMC method for new solid-state qubit systems where experimental or theoretical reference data is not available. A straightforward way out of this problem would be to only use S.R. orbitals for representing all the states, but at least in this study this led to larger errors in the excitation energy of the $3E$ state. Further improvements to the wave functions can be obtained with new orbital optimization schemes or larger multideterminant expansions, which could also resolve the aforementioned problems.

It should be noted that while larger multideterminant expansions could resolve many issues here, it is extremely hard to extrapolate simultaneously to infinite system size and full CI limit, and the improvement of the nodal surfaces with larger determinantal expansions will only work if the nodal errors decay quickly with the number of determinants.  

After resolving the issue of overestimation in the excited state energies, many interesting studies may follow. The use of fully correlated QMC can allow new, more complicated many-body studies of the magneto-optical properties of spin qubits. For example, the effects of surfaces, interfaces, or other defects to the excitation energies are difficult to simulate with cluster-based approaches or embedding schemes, but would be accessible with the $>200$ atom simulation cells used in this study. 

Many properties relevant for spin qubits require further development with QMC methods. For example, the study of structural properties and potential energy surfaces with QMC would be very useful. Studies on the estimation of atomic forces with QMC \cite{rios2019,yang2023} and structural optimization with stochastic forces \cite{chen2022} already exist. Another issue is the simulation of electron-phonon interactions, that play a key role in non-radiative transitions \cite{goldman2015b}. The phononic effects can be estimated by sampling atomic configurations based on the vibrational states estimated at DFT level, but DFT cannot treat the singlets.

Also the understanding of spin-orbit interactions and zero-field splittings are very important in harnessing the utility of the NV$^{-}$-center and related point defects. Studying observables dependent on spin would require QMC to include spin degree of freedom to the wave funtions, as has been done in a recent implementation \cite{melton2016}.  

The structural relaxation with QMC would be expensive, and certain DFT studies show that estimation of spin-related quantities would require simulation cells consisting of up to thousand atoms \cite{thiering2017}.

The development of embedding schemes for QMC are very possibly required in order to perform in-depth studies of spin defects. Embedding schemes allowing large multideterminant expansions with QMC already exist \cite{christlmaier2022}. DMC embedding could be done in much larger volumes or orbital spaces than with the current quantum chemistry embedding approaches. This would reduce the errors introduced by current embedding schemes originating from coupling of the states from embedding region to the bulk material.

\begin{acknowledgments}
We would like to thank Dr. Neil Drummond for useful discussions and insights related to QMC simulation of defects.  We acknowledge the generous computational resources provided by CSC (Finnish IT Centre for Science), in particular to Mahti and Lumi clusters.  This work was partially supported by the Academy of Finland Grants No. 285809, No. 293932, No. 319178, No. 334706, and No. 334707.
\end{acknowledgments}

\appendix

\section{}

\subsection{The Hamiltonian and Symmetry Groups}

The Hamiltonians $\hat{H}$,$\hat{H}_0$, and $\hat{H}^i_0$, as presented in Eq. (\ref{equation: interacting Hamiltonian}), are invariant under operations of their corresponding symmetry groups $\mathbb{G}(\hat{H})$,$\mathbb{G}(\hat{H}_0)$, and $\mathbb{G}(\hat{H}_0^i)$. The symmetry group of both $\hat{H}$ and $\hat{H}_0^i$ is $C_{3v}$. Each Hamiltonian and operations of its symmetry group commute and have common eigenstates. 

Any symmetry operation $T_g$ of the group member $g$ of the symmetry group $\mathbb{G}(\hat{H}^i_0)$ can be represented as a matrix acting in a $3$-dimensional coordinate space. The choice of appropriate matrices for all $g\in \mathbb{G}(\hat{H}^i_0)$ defines a representation $\Gamma_{\mathbb{G}\left(\hat{H}^i_0\right)}$ that maps the symmetry operation $T_g$ into a matrix $\Gamma_{\mathbb{G}\left(\hat{H}^i_0\right)}(g)$. 

As $\hat{H}_0=\sum_i\hat{H}_0^i$, the representations of $\mathbb{G}(\hat{H}_0)$ are tensor products of the representations of $\mathbb{G}(\hat{H}^i_0)$:
\begin{equation}
  \label{equation: many-body tensor product symmetry}
  \Gamma_{\mathbb{G}\left(\hat{H}_0\right)}=\Gamma_{\mathbb{G}\left(\hat{H}_0^1\right)} \otimes \Gamma_{\mathbb{G}\left(\hat{H}_0^2\right)} \otimes ... \otimes \Gamma_{\mathbb{G}\left(\hat{H}_0^N\right)},
\end{equation} 
Each particle in the non-interacting picture experience the same set of symmetries $\mathbb{G}\left(\hat{H}_0^i\right)=\mathbb{G}\left(\hat{H}_0^j\right) \forall i,j$. 

The symmetry group of the interacting Hamiltonian $\mathbb{G}(\hat{H})$ has also representations of the form in Eq. (\ref{equation: many-body tensor product symmetry}), but the many-body symmetry operations $T_{g_1,g_2,...,g_N}=T_{g_1}T_{g_2}...T_{g_N}$, with $g_i$ referring to operation in particle $i$, are restricted: $T_{g_1}=T_{g_2}=...=T_{g_N}$. This is because the interparticle interactions described by $\hat{H}_2$ are invariant under rotation of one particle only if all the other particles are rotated equivalently. Hence $\mathbb{G}(\hat{H})$ is isomorphic to $\mathbb{G}(\hat{H}_0^i)$ \cite{towler2000}, and both are subgroups of $\mathbb{G}(\hat{H}_0)$. 

\subsection{Single-particle states}

The many-body states can be derived using the single-particle states with the molecular orbital method \cite{coulson1957}, as has been done in numerous works before \cite{lenef1996,maze2011}. However, we repeat the process here as our basis of Slater determinants decomposed into spin up- and down-components gives rise to different forms of the wavefunctions than previously.

The single-particle states localized around the NV$^{-}$-center are treated in the basis of four dangling bonds $\sigma_i$ created in the removal of a carbon atom from the perfect diamond lattice. Dangling bonds $\sigma_1,\sigma_2$, and $\sigma_3$ are attached to the carbon atoms neighbouring the vacancy, and $\sigma_N$ to the nitrogen impurity. The dangling bonds constitue a basis for the Hilbert space of $\hat{H}_0^i$. 

By applying a group theoretic projection operator,
\begin{equation}
  \label{equation: projection operator}
  \mathbf{P}^{\alpha}=\frac{n_{\alpha}}{\lambda_{\mathbb{G}(\hat{H}_0^i)}}\sum_{g \in \mathbb{G}(\hat{H}_0^i)}\chi^{(\alpha)*}(g)\Gamma_{\mathbb{G}(\hat{H}_0^i)}(g)
\end{equation}
to $\sigma_i$ we can find the orbitals transforming according to the irreducible representations $\alpha$ of $\mathbb{G}(\hat{H}_0^i)$, i.e. $C_{3v}$. $\alpha=\{ A1, A2, E \}$, meaning that $C_{3v}$ consists of two completely symmetric ($A_1,A_2$) and one doubly degenerate ($E$) irreducible representations. Above, $n_\alpha$ is the dimensionality of the irreducible representation $\alpha$, $\lambda_{\mathbb{G}(\hat{H}_0^i)}$ the order of the group $\mathbb{G}(\hat{H}_0^i)$, and $\chi^\alpha (g)$ is the character of the group member $g$ under the irreducible symmetry representation $\Gamma^\alpha_{\mathbb{G}(\hat{H}_0^i)}$. The orbitals are \cite{maze2011}
\begin{align}
  \label{equation: sp states}
  \begin{aligned}
    a_N&=\sigma_4 \\
    a_1&= \left[ \ \sigma_1+\sigma_2+\sigma_3 \right]/3\\
		e_x&= \left[ 2\sigma_1-\sigma_2-\sigma_3 \right]/\sqrt{6} \\
		e_y&= \left[ \ \ \ \ \ \ \ \ \sigma_2-\sigma_3 \right]/\sqrt{2}.
  \end{aligned}
\end{align}
$a_N$ and $a_1$ are completely symmetric and transform accoring to $\Gamma^{A_1}_{\mathbb{G}(\hat{H}_0^i)}$, and $e_x$ and $e_y$ are degenerate orbitals transforming as $\Gamma^{E}_{\mathbb{G}(\hat{H}_0^i)}$. $a_N$ is energetically below the valence band, and is assumed to be filled by $2$ electrons. $a_1$, $e_x$ and $e_y$ are depicted in Fig. \ref{figure: orbital depictions}, and are within the diamond band gap, as shown in Fig. \ref{figure: orbital bands}. 

\subsection{Many-body states}

In the Slater-Jastrow type wave functions (see Eq. (\ref{equation:slater-jastrow wf, qualitative})), the Jastrow factor $J(\mathbf{R})$ is fully symmetric and the Slater determinants $D_n=D_n^\uparrow D_n^\downarrow$ consists of $N^\uparrow$ spin-up and $N^\downarrow$ spin-down electrons filling the single-particle orbitals. The determinants transform as basis functions of the irreducible representations of $\mathbb{G}(\hat{H}_0)$. As $\mathbb{G}(\hat{H}) \subset \mathbb{G}(\hat{H}_0)$, we can decompose this irreducible representation according to irreducible representations of $\mathbb{G}(\hat{H})$ as
\begin{equation}
  \label{equation: general decomposition of H0 to H states}
  \Gamma_{\mathbb{G}(\hat{H}_0)}=\Gamma^1_{\mathbb{G}(\hat{H})} \otimes \Gamma^2_{\mathbb{G}(\hat{H})} \otimes ... \otimes \Gamma^M_{\mathbb{G}(\hat{H})},  
\end{equation}
based on the character system of the representations. Above, the irreducible representations $\Gamma^\alpha_{\mathbb{G}(\hat{H})}$ are said to be compatible with $\Gamma_{\mathbb{G}(\hat{H}_0)}$.

For the symmetry analysis of the many-body states, it is sufficient to only consider the open shell of the states, i.e. electrons in the $a_1$ and $e_{x,y}$ states. Instead of considering the states filled by electrons we can consider the holes in the open shell to simplify the analysis. In the ground state occupation, $2$ electrons fill the $a_1$ state and $2$ electrons occupy the degenerate $e$ states. Thus we have $2$ holes of $E$ symmetry. The optical excitation in the single-particle picture corresponds to promotion of an electron from the $a_1$ to one of the degenerate $e$ states, and hence there are $2$ holes of symmetries $A$ and $E$. The symmetry representations $\Gamma_{\mathbb{G}(\hat{H}_0)}$ for these two occupations can be decomposed accoring to Eq. (\ref{equation: general decomposition of H0 to H states}) as
\begin{align}
  \begin{aligned}
    \Gamma^E_{\mathbb{G}(\hat{H}_0)} \otimes \Gamma^E_{\mathbb{G}(\hat{H}_0)}
    =& \Gamma^{A_1}_{\mathbb{G}(\hat{H})} \otimes \Gamma^{A_2}_{\mathbb{G}(\hat{H})}
    \otimes \Gamma^E_{\mathbb{G}(\hat{H})} \\
    \Gamma^{A_1}_{\mathbb{G}(\hat{H}_0)} \otimes
    \Gamma^E_{\mathbb{G}(\hat{H}_0)} =&
    \Gamma^E_{\mathbb{G}(\hat{H})}
  \end{aligned}
\end{align}

To find the eigenstates of $\hat{H}$, we apply the projector operator of Eq. (\ref{equation: projection operator}), but with a representation $\Gamma_{\mathbb{G}(\hat{H})}$, to all the possible Slater determinants formed by populating the orbitals $a_1$, $e_x$, and $e_y$ of both spin channels with $4$ electrons. Table \ref{table: many-body wave functions} shows the many-body wavefunctions, without the Jastrow factor for clarity, obtained with the projector method.

\subsection{Variational principle of the many-body states}

In VMC the variational principle holds for excited states that are orthogonal to the lower states, for example in cases where the excitation represents a different irreducible representation of the symmetry group of the system than the lower states. In fixed-node DMC the orthogonality of the excited state against the ground state is not enough to quarantee the variational principle \cite{foulkes1999}. Only states both orthogonal to the ground state and transforming according to one-dimensional irreducible representations of the symmetry group of the system have a variational principle. This means that the variational principle applies only to the $1A_1$ state.

Because both $1E$ and $3E$ transform according to a $2$-dimensional irreducible representation, they may collapse to lower states during a DMC simulation. However, as the spin is conserved during the simulations, we can close out the possibility of a spin-up (or -down) projection of the $3E$ state to collapse to $1E$-state. There is still the possibility of any of the spin projections of the $3E$-state to collapse to the $3A_2$-state. Also the singlet $1E$-state can collapse to the zero-spin projection of the $3A_2$ state.

According to Foulkes \textit{et al.} \cite{foulkes1999}, we can establish a weaker, or even the original strong variational principle for states transforming as multidimensional irreducible representations under certain conditions. This is done by considering the subgroups of $\mathbb{G}(\hat{H})$, and finding the compatible irreducible representations of the subgroups. If there is a subgroup that has one-dimensional irreducible representations compatible with the excited state symmetry, but not compatible with the ground state, a weaker variational principle holds.  

The irreducible representation E of $C_{3v}$ can be decomposed with respect to irreducible representations of its subgroup $C_s$ as
\begin{equation}
  \Gamma^E_{C_{3v}} =  \Gamma^{A_1}_{C_s} \otimes \Gamma^{A_2}_{C_s}
\end{equation}
Out of the irreducible representations $A_1$ and $A_2$ of $C_s$, $A_1$ is not compatible with the ground state symmetry of the NV$^-$-center, $A2$. This means that both the $3E$ and $1E$ states have a weaker variational principle stating that the DMC energy of the states is greater than or equal to the many-body state transforming according to the irreducible representation $\Gamma^{A_1}_{C_s}$. Using the projection operator
\begin{equation}
  \label{equation: cs projection operator}
  P^{A_1}=\frac{n_{A_1}}{\lambda_{C_s}}\sum_{g \in C_s} \chi^{A_1}(g)\Gamma^{A_1}_{C_s}(g),
\end{equation}
we can project this state out of the spin-up projection of $3E$ and $1E$:
\begin{align}
  \label{equation: cs projections}
  \begin{aligned}
    P^{A_1}\Psi^{3E}_{x,1}&=0 \\
    P^{A_1}\Psi^{3E}_{y,1}&=\Psi^{3E}_{y,1} \\
    P^{A_1}\Psi^{1E}_x&=0 \\
    P^{A_1}\Psi^{1E}_y&=\Psi^{1E}_y
  \end{aligned}
\end{align}
Equation (\ref{equation: cs projections}) shows that the $y$-components of the spin-up state of $3E$ and $1E$ already transform according to $\Gamma^{A_1}_{C_s}$. Therefore the variational principle of $\Gamma^{A_1}_{C_s}$ is also the variational principle of $\Gamma^{E}_{\hat{H}}$. Since spin-up component of $3E$ cannot collapse into the singlet states during DMC it is the lowest spin-up state that $\Gamma^{A_1}_{C_s}$ is compatible with. $1E$ is the lowest spin-zero state that $\Gamma^{A_1}_{C_s}$ is compatible with. Therefore we see that the strong variational principle applies for the $y$-components of both spin-up (and -down) $3E$ and $1E$.

%\bibliographystyle{unsrt}
%\bibliography{article.bib}

\end{document}